\documentclass[a4paper]{jpconf}
\usepackage{graphicx}
\begin{document}
\title{An ERC Starting Grant project on p-process nucleosynthesis concluded}

\author{Gy Gy\"urky, Z Hal\'asz, T Sz\"ucs, G G Kiss, Zs F\"ul\"op}

\address{Institute for Nuclear Research (MTA Atomki), H-4001 Debrecen, POB.51., Hungary}
\ead{gyurky@atomki.mta.hu (Gy.\,Gy\"urky)}

\begin{abstract}

In 2008 a Starting Grant project supported by the European Research Council titled ``\textit{Nuclear reaction studies relevant to the astrophysical p-process nucleosynthesis}'' was launched. After five years of successful research related to the experimental investigation of proton- and alpha-induced nuclear reaction for the astrophysical p-process, the project came to an end. In this paper a summary of the research and the most important achievements is given.

\end{abstract}

\section{Introduction}

Nucleosynthesis, the production of chemical elements that build up our universe is an extremely complicated scientific problem. Not surprisingly, a whole new interdisciplinary science called nuclear astrophysics was born at around the middle of the 20th century to devise and study the processes leading to the production of various groups of chemical elements \cite{B2FH}. After several decades of progress, we have now a reasonably good general understanding of the astrophysical sites and processes contributing to the nucleosynthesis. Our models are, however, still very far away from being able to describe well the different astrophysical phenomena and the composition of the universe, so the quest for the origin of the elements is still unfinished \cite{jos11}.

The chemical elements heavier than iron represent a special group in nuclear astrophysics. Since their synthesis requires rather than produces energy, their production cannot be the source of energy that powers a star, their existence in nature must be attributed to some kind of a  subsidiary process. Since the capture of neutrons is not hindered by the Coulomb-barrier, the heavy elements are thought to be produced mainly through neutron capture reactions. Two distinct processes of neutron capture, the s- and r-processes play a role in the heavy element synthesis which takes place in different stellar environments \cite{kap11,arn07}.

About three dozens of heavy stable isotopes located on the proton rich side of the valley of nuclear stability, on the other hand, cannot be produced by any of these two neutron capture processes. These are the so-called p-isotopes and their stellar production mechanism is generally referred to as the astrophysical p-process \cite{woo78}. The general feature of the p-isotopes is their typically very low elemental abundance (with a few exceptions below 1\,\%). As opposed to the s- and r-isotopes, no heavy element has a dominant p-isotope. Therefore, abundance information of the p-isotopes is available only from the Solar System (from meteorites or from the bulk), the performance of telescopes does not allow the identification of the p-isotopes in stellar spectra. 

Therefore, p-process models try to reproduce the Solar System p-isotope abundances. However, in spite of the significant progress huge differences between the observed and calculated p-isotope abundances persist \cite{dil08}. In order to cure the discrepancies, several different processes are considered which might contribute to the p-isotope production. Indeed, the p-process is actually thought to comprise several sub-processes such as the $\nu$-, $\nu$p-, rp-, pn-processes \cite[e.g.]{woo90,fro06,sch98,gor02}. None of these processes can, however, be responsible for the production of the p-isotopes in the whole mass range. Only the so-called $\gamma$-process is able to produce proton-rich isotopes in the whole mass range and thus the $\gamma$-process seems to be by far the most important sub-process of the p-process.

The $\gamma$-process proceeds through $\gamma$-induced reactions initiated on pre-existing heavy isotopes. In a high temperature stellar environment the high energy wing of the Planck distribution is energetic enough to remove neutrons from heavy isotopes. Consecutive ($\gamma$,n) reactions drive thus the material towards the proton-rich region producing this way the p-isotopes or their radioactive progenitors. As the isotopes become more and more neutron-deficient, charged particle emitting ($\gamma$,p) and ($\gamma,\alpha$) reactions start to play a role and influence strongly the resulting p-isotope abundances. As the high temperature is a necessary prerequisite of the $\gamma$-induced reactions, explosive scenarios are thought to host the $\gamma$-process. Perhaps the most likely and most intensively studied site is the O/Ne layer of core collapse supernovae \cite{ray95}, but type Ia supernovae are also considered and being studied \cite{tra11}.

\section{Data needs for the $\gamma$-process modelling}

For a successful $\gamma$-process modelling the astrophysical conditions must of course be known. This involves the knowledge about the seed isotope abundances and the conditions of the explosion such as the spatial and temporal variation of density and temperature. Equally important is the nuclear physics input. A full $\gamma$-process network calculation involves thousands of reactions, the rate of which must be known for a reliable calculation. Most of the reactions take place on radioactive nuclei, therefore it is not surprising that experimental data are very scarce and the models have to rely on theoretical reaction rates obtained typically from the Hauser-Feshbach statistical model. The fact that the $\gamma$-process models fail to reproduce the observed p-isotope abundances can be related to the possibly incorrect reaction rates from theory. Therefore, the experimental study of $\gamma$-process related reactions is of high importance in order to provide direct data for the $\gamma$-process networks and to check the reliability of statistical model calculations.

The $\gamma$-induced reactions which play the key role in the $\gamma$-process can be studied directly if a suitable photon beam is available. However, it is technically much easier to study the inverse capture reactions and to obtain the $\gamma$-induced reaction rate by using detailed balance. Moreover, the study of capture reactions gives a more direct information for the astrophysical process, since the ground state contribution (which is always measured in the laboratory) is much larger in the case of capture reactions than for $\gamma$-induced ones. The measurement of capture cross sections in the relevant mass and energy range can thus provide important input for the $\gamma$-process models and can contribute to the better understanding of the origin of the p-isotopes.

Owing to their high importance in several different fields of science and technology, neutron capture cross sections of stable isotopes are extensively studied and relatively well known. In the case of charged particle capture reactions, however, the experimental database relevant for the $\gamma$-process is very limited \cite{KADONIS}. The reason for this is that in the astrophysically relevant energy range the cross sections are typically tiny and thus practically no experimental effort was devoted to these reactions before the urgent need from the $\gamma$-process modelling arose. Therefore, the statistical models in the case of charged particle induced reactions are largely untested. On the other hand, dedicated sensitivity studies indicate that the resulting p-isotope abundances from a network calculation depend strongly on the reaction rates of the ($\gamma$,p) and ($\gamma,\alpha$) reactions \cite{rau06,rap06}. It is therefore necessary to extend the available experimental database for charged particle induced reactions providing this way a stringent test of the statistical model calculations and also the direct input for the $\gamma$-process networks.

Answering the call from astrophysics modelers, a systematic study of charged particle induced reactions relevant for the $\gamma$-process was started about 15 years ago \cite{som98}. Significant part of the experimental database was created by the nuclear astrophysics group of Atomki. The Starting Grant of the European Research Council in the last five years (2008-2013) made a very important contribution to the success of the experiments. Some details of the project and the obtained results are presented in the next sections.

\section{Alpha-induced reaction cross section measurements}

The above mentioned sensitivity studies indicate that the calculated p-isotope abundances are especially sensitive to the rates of the ($\gamma,\alpha$) reactions in the heavier mass range of the p-isotopes. This mass range was unfortunately completely unexplored experimentally. With the exception of $^{144}$Sm no ($\alpha,\gamma$) cross section measurement above A\,$\approx$\,120 was available. Therefore, one of the main aims of the project was to extend the available experimental database towards the heavy mass region. 

Table \ref{tab:ag} shows the list of $\alpha$-induced reactions which were studied in the framework of the ERC project. The experiments were always carried out at the cyclotron accelerator of Atomki using the activation method. The reader is referred to the original publications for further experimental details.

\begin{table}
	\centering
	\caption{\label{tab:ag} List of $\alpha$-induced reactions studied in this project. The quoted energies are given in the center-of-mass frame. The study of several other reactions not listed in this table is in progress.}
		\begin{tabular}{ccc|ccc}
\hline
Reaction & Energy range & Ref. & Reaction & Energy range & Ref.\\
 & [MeV] & & & [MeV] \\
\hline
$^{64}$Zn($\alpha,\gamma$)$^{68}$Ge & 8.0 -- 12.4 & \cite{gyu12,gyu12c} & $^{130}$Ba($\alpha,\gamma$)$^{134}$Ce & 11.6 -- 16.0 & \cite{gyu11,hal12b} \\
$^{64}$Zn($\alpha$,n)$^{67}$Ge & 9.0 -- 12.4 & \cite{gyu12,gyu12c} & $^{130}$Ba($\alpha$,n)$^{133}$Ce & 12.1 -- 16.0 & \cite{gyu11,hal12b}\\
$^{64}$Zn($\alpha$,p)$^{67}$Ga & 5.8 -- 12.4 & \cite{gyu12,gyu12c} & $^{132}$Ba($\alpha$,n)$^{135}$Ce & 12.1 -- 16.0 & \cite{gyu11,hal12b} \\			
$^{113}$In($\alpha,\gamma$)$^{117}$Sb & 8.7 -- 13.6 & \cite{yal09,gur09,yal09b} & $^{151}$Eu($\alpha,\gamma$)$^{155}$Tb & 12.3 -- 15.6 & \cite{gyu10,gyu10b}\\
$^{113}$In($\alpha$,n)$^{116}$Sb & 9.7 -- 13.6 & \cite{yal09,gur09,yal09b} & $^{151}$Eu($\alpha$,n)$^{154}$Tb & 11.3 -- 15.6 & \cite{gyu10,gyu10b}\\
$^{127}$I($\alpha,\gamma$)$^{131}$Cs & 9.5 -- 15.2 & \cite{kis12b} & $^{169}$Tm($\alpha,\gamma$)$^{173}$Lu & 12.3 -- 17.1 & \cite{kis11,kis11b,rau12}\\
$^{127}$I($\alpha$,n)$^{130}$Cs & 9.6 -- 15.2 & \cite{kis12b} & $^{169}$Tm($\alpha$,n)$^{172}$Lu & 11.2 -- 17.1 & \cite{kis11,kis11b,rau12} \\
\hline
		\end{tabular}
\end{table}

The obtained cross sections were always compared with the predictions of statistical model calculations. For this purpose the NON-SMOKER code \cite{NONSMOKER} was typically used since reactions rates provided by this code are often used in astrophysical network calculations. Other codes like TALYS \cite{TALYS} or the recently developed SMARAGD \cite{SMARAGD} were also used occasionally in order to assess the influence of statistical model implementation and the default input parameters of the codes on the resulting cross sections. 

The general observation is that the ($\alpha,\gamma$) cross sections calculated using the default settings of the NON-SMOKER code always overestimate significantly the measured data. The deviation is between a factor of two and almost one order of magnitude. In order to draw an astrophysical conclusion from this observation, however, two things have to be considered. First, the results of the statistical model calculations depend on a few input parameters, such as e.g. the optical potentials, level densities, $\gamma$-ray strength functions, etc. We do not have a priori information about which of the several possible choices of these input parameters are correct. The comparison with experiment can help choose the best parameters. Second, the studied energy range of the ($\alpha,\gamma$) reactions is unfortunately higher than the astrophysically relevant one. The experimental energy range is limited by the falling cross section towards low energies. Therefore, the comparison of theory and experiment can be done only at energies without direct astrophysical relevance. 

Unfortunately the calculated cross sections show a different sensitivity to various input parameters at different energy regions. For a detailed study of these sensitivities see Ref. \cite{rau12b}. At the experimentally accessible energy range the calculations show high sensitivity to more than one input parameter, the cross sections are typically sensitive to the $\alpha$-, $\gamma$- and neutron widths. Therefore it is not possible to constrain one of these input parameters based on the data. Studying a different reaction channel can, however, help. Although not directly relevant astrophysically, the ($\alpha$,n) reaction channel can be used to constrain the $\alpha$-width since the calculated ($\alpha$,n) cross section is typically sensitive only to this width. Fortunately, in those cases where the ($\alpha,\gamma$) reaction leads to a radioactive isotope, the final nucleus of the ($\alpha$,n) channel is often also radioactive. Using the activation method the simultaneous determination of the ($\alpha,\gamma$) and ($\alpha$,n) cross sections is thus possible. Therefore, along with the ($\alpha,\gamma$) cross section measurement, the ($\alpha$,n) channel was also studied, as indicated in table \ref{tab:ag}. 

The $\alpha$-width which is obtained from the $\alpha$-nucleus optical potential, turns out to be perhaps the most important input parameter of the calculations for reactions involving $\alpha$-particles. Different global $\alpha$-nucleus optical potentials lead to largely different calculated cross section. See e.g. figure 3. in ref.\,\cite{rau12}. This introduces a high uncertainty in $\gamma$-process network calculations and therefore the study of the low energy $\alpha$-nucleus optical potential is of high importance. The presently available experimental dataset of ($\alpha,\gamma$) reactions does not show a clear picture yet about which global $\alpha$-nucleus optical potential is the best suited for the  $\gamma$-process network calculations. Further $\alpha$-induced cross section data is therefore clearly needed.

\section{Alpha elastic scattering experiments}

The important $\alpha$-nucleus optical potential can be directly studied with high precision $\alpha$-elastic scattering experiments. Such experiments were the other main objective of the present ERC project. Table \ref{tab:aa} shows those scattering reactions which were studied in the framework of the present project.

\begin{table}
	\centering
	\caption{\label{tab:aa} List of $\alpha$-elastic scattering reactions studied in this project. The quoted energies are given in the center-of-mass frame. The study of several Sm and Nd isotopes is in progress.}
		\begin{tabular}{ccc}
\hline
Reaction & Energy [MeV] & Ref. \\
\hline
$^{64}$Zn($\alpha,\alpha$)$^{64}$Zn & 11.3, 15.2 & * \\
$^{89}$Y($\alpha,\alpha$)$^{89}$Y & 15.5, 18.6 & \cite{kis09} \\
$^{110}$Cd($\alpha,\alpha$)$^{110}$Cd & 15.6, 18.9 & \cite{kis11c,kis12} \\
$^{116}$Cd($\alpha,\alpha$)$^{116}$Cd & 15.6, 18.9 & \cite{kis11c,kis12} \\
$^{113}$In($\alpha,\alpha$)$^{113}$In & 15.6, 18.8 & \cite{kis13} \\
$^{115}$In($\alpha,\alpha$)$^{115}$In & 15.6, 18.8 & \cite{kis13b}* \\
\hline
* {\footnotesize Analysis in progress}
		\end{tabular}
\end{table}

The angular distribution of elastically scattered $\alpha$-particles were measured in a full angular range between typically 20 and 175 degrees. The chosen energies represent a compromise between the astrophysically relevant energies and energies where the scattering cross section deviates significantly from the Rutherford cross section. The experiments were thus carried out at around the Coulomb-barrier. The angular distributions were measured at a high precision scattering chamber mounted on one of the beamlines of the cyclotron accelerator of Atomki.

The measured angular distributions were compared with the predictions using several different global $\alpha$-nucleus optical potentials. The general observation is that none of the optical potentials can give a good description of the measured data and they fail to reproduce the evolution of the elastic scattering cross section along isotopic and isotonic chains. The best optical potentials can still be selected based on the scattering data, but unfortunately it is found that those potentials which give the best description of e.g. the ($\alpha,\gamma$) cross sections, fail completely to reproduce the scattering data. This also points to the need of an improved low energy $\alpha$-nucleus optical potential parametrization for astrophysical purposes.

Based on the scattering data obtained in this project and previously at Atomki, we have developed a new, global optical potential \cite{moh13}. The aim of this new potential is to provide input for astrophysical applications, i.e. the potential focuses on low energies. The potential has only five parameters, so the extrapolation towards unstable isotopes is hoped to be more reliable than in the case of a many parameter global potential. The new potential predicts well the total cross section of $\alpha$-induced reactions and provides a relatively good reproduction of the available scattering data. Further development of this potential is in progress with the inclusion of additional scattering data and fine tuning based on $\alpha$-induced reaction data.

\section{Proton-induced reactions and auxiliary measurements}

Realizing their importance and the lack of experimental data, the main focus of the ERC project was the study of $\alpha$-induced reactions and $\alpha$-elastic scattering. Nevertheless, some proton-induced reactions relevant for the $\gamma$-process were also studied. 
(p,$\gamma$) and (p,n) cross sections of several stable Se isotopes was measured in the Gamow window relevant for the $\gamma$-process \cite{ska11}. The standard NON-SMOKER calculations give a reasonably good description of the measured data. Through the study of the $^{85}$Rb(p,n)$^{85}$Sr reaction we demonstrated the Coulomb suppression of the stellar enhancement factor, i.e. that the stellar effects can be minimized in the charged particle channel, even when the reaction Q value is negative \cite{kis08,rau09}. For the $^{92}$Mo(p,$\gamma$)$^{93}$Tc reaction ambiguous cross section data can be found in the literature. In order to study the discrepancy, this reaction was studied with a method never applied for any $\gamma$-process related reaction so far. The cross section is derived from thick target yield measurement with activation \cite{gyu14}. 

The cross sections of charged particle induced reactions were determined with the activation method in this project. The uncertainty of the half-life of the reaction products influences therefore directly the accuracy of the measured cross sections. In some cases where the half-lives in question were not known with sufficient precision, dedicated precise half-life measurement was carried out. The half-life of an isomer in $^{154}$Tb was determined to be 9.994\,$\pm$\,0.039\,h, a value which is one order of magnitude more precise than the literature value \cite{gyu09}. The half-life of $^{133m}$Ce was measured to be 5.326\,$\pm$\,0.011\,h, a factor of 30 more precise value than the literature one \cite{far11}. The half-life of $^{66}$Ga, an isotope important for high energy efficiency calibration of $\gamma$-detectors, was measured to be 9.312\,$\pm$\,0.032\,h, supporting the validity of one of the two recent, strongly contradicting measurements \cite{gyu12b}.

One of the most important methodological development of the project was the extension of the activation method by the detection of characteristic X-ray radiation. In many cases, especially in the higher mass region of p-isotopes the decay of the radioactive reaction products is not followed by a high intensity $\gamma$-radiation. The conventional $\gamma$-counting based activation is therefore not applicable. These isotopes, on the other hand, typically decay by electron capture which is followed by characteristic X-ray emission. The detection of this radiation can be used for the cross section determination. The applicability of the method using a LEPS (Low Energy Photon Spectrometer) detector was demonstrated with the $^{169}$Tm($\alpha,\gamma$)$^{173}$Lu reaction \cite{kis11b} and is used and will be used for several other reactions.

\section{Conclusions and outlook}

During the five years of the project charged particle induced reactions and elastic scattering reactions relevant for the astrophysical $\gamma$-process were extensively studied. The experimental database relevant for this poorly known nucleosynthetic process was substantially increased. Our present knowledge about the origin of p-isotopes has been summarized recently in a review paper \cite{rau13}, where the experimental situation of the nuclear physics side of $\gamma$-process is also surveyed. 

The number of studied reactions, however, still remains very low compared to the thousands of reactions involved in a $\gamma$-process reaction network. And although great progress was made in astrophysical modeling recently, the models are still far from being able to reproduce the observed p-isotope abundances. Therefore, the experimental study of $\gamma$-process relevant reactions cannot stop. 

The study of several other reactions not listed in tables \ref{tab:ag} and \ref{tab:aa} is in progress at Atomki\footnote{Since the submission of this manuscript the results of some of these measurements have been published. The list below is updated with these references.}. These include among others: $^{121}$Sb($\alpha$,$\gamma$)$^{125}$I, $^{121}$Sb($\alpha$,n)$^{124}$I and $^{123}$Sb($\alpha$,n)$^{126}$I \cite{kor13}; $^{107}$Ag($\alpha,\gamma)^{111}$In and $^{107}$Ag($\alpha,n)^{110}$In \cite{yal13}; $^{115}$In($\alpha$,$\gamma$)$^{119}$Sb, $^{162}$Er($\alpha$,$\gamma$)$^{166}$Yb, $^{162}$Er($\alpha$,n)$^{165}$Yb \cite{kis14}, $^{164}$Er($\alpha$,n)$^{167}$Yb, $^{166}$Er($\alpha$,n)$^{169}$Yb \cite{kis15}, $^{191}$Ir($\alpha$,$\gamma$)$^{195}$Au, $^{191}$Ir($\alpha$,n)$^{194}$Au and $^{193}$Ir($\alpha$,n)$^{196}$Au \cite{szu13}; and elastic scattering on Sm and Nd isotopes. 

The experimental methods have to be developed, however, as more and more difficult-to-study reactions are faced. The so far very fruitful activation method can still be extended and further reactions can be investigated with this technique. For example, the amount of produced long-lived reaction products can be determined by the AMS (Accelerator Mass Spectrometry) technique. Pilot measurements with this technique is in progress for the $^{64}$Zn($\alpha,\gamma$)$^{68}$Ge and $^{142}$Nd($\alpha,\gamma$)$^{146}$Sm reactions.

In those cases where the reaction leads to a stable product, the activation method cannot be applied and the cross section must be determined with the technically much more challenging in-beam method. Such measurements are in progress in several institutions with different experimental approaches \cite[e.g.]{sau11,hal12,har13,sim13}.

The actual $\gamma$-process path is located on the proton rich side of the valley of stability mainly in the region of radioactive isotopes. Thus most of the reactions relevant for a $\gamma$-process network cannot be studied with a stable target experiment. In order to be able to measure capture cross sections on radioactive isotopes, the rapidly developing field of radioactive ion beams needs to be exploited. A pilot experiment at the Experimental Storage Ring of GSI on the $^{96}$Ru(p,$\gamma$)$^{97}$Rh reaction demonstrated the feasibility of such a measurement \cite{zho10} and we hope that the measurements will be extended towards the radioactive proton-rich isotopes in the near future.

\section*{Acknowledgments}
This work would not have been possible without the valuable support from the European Research Council Starting Grant No. 203175. Part of the research was also supported by OTKA grants K101328, K108459, PD104664 and NN83261(EuroGENESIS). G G Kiss is a Bolyai fellow. Besides the present nuclear astrophysics group of Atomki signing this paper, many other people contributed to the success of the project. Therefore, we thank J. Farkas, D. Galaviz, Z. Korkulu, P. Mohr, A. Ornelas, E. Somorjai, P.M. Tak\'acs, Zs. T\"or\"ok, C. Yal\c c\i n and many others for their work. The theoretical analysis and interpretation of the measured data was always done by T. Rauscher.

\end{document}